\documentclass[aps,showpacs,amsmath,amssymb,superscriptaddress,prl,preprint]{revtex4-1}

\usepackage{graphicx}
\usepackage{bm}
\usepackage{color}
\usepackage{epstopdf}
\usepackage[pagewise]{lineno}

\begin{document}

\title{Impact of inherent energy barrier on spin-orbit torques in magnetic-metal/semimetal heterojunctions}

\author{Tenghua Gao\footnote{These authors contributed equally\label{a}}}
\affiliation{Keio Institute of Pure and Applied Science, Keio University, Yokohama 223-8522, Japan}
\affiliation{Department of Applied Physics and Physico-Informatics, Keio University, Yokohama 223-8522, Japan}
\affiliation{State Key Laboratory of Advanced Technology for Materials Synthesis and Processing, Wuhan University of Technology, Wuhan, China}

\author{Alireza Qaiumzadeh\textsuperscript{\ref{a}}}
\affiliation{Center for Quantum Spintronics, Department of Physics, Norwegian University of Science and Technology, NO-7491 Trondheim, Norway}

\author{Roberto E. Troncoso\textsuperscript{\ref{a}}}
\affiliation{Center for Quantum Spintronics, Department of Physics, Norwegian University of Science and Technology, NO-7491 Trondheim, Norway}
\affiliation{School of Engineering and Sciences, Universidad Adolfo Ib\'a\~nez, Santiago, Chile}

\author{Satoshi Haku}
\affiliation{Department of Applied Physics and Physico-Informatics, Keio University, Yokohama 223-8522, Japan}

\author{Hongyu An}
\affiliation{College of New Materials and New Energies, Shenzhen Technology University, Shenzhen 518118, China}

\author{Hiroki Nakayama}
\affiliation{Department of Applied Physics and Physico-Informatics, Keio University, Yokohama 223-8522, Japan}

\author{Yuya Tazaki}
\affiliation{Department of Applied Physics and Physico-Informatics, Keio University, Yokohama 223-8522, Japan}

\author{Song Zhang}
\affiliation{State Key Laboratory of Advanced Technology for Materials Synthesis and Processing, Wuhan University of Technology, Wuhan, China}

\author{Rong Tu}
\affiliation{State Key Laboratory of Advanced Technology for Materials Synthesis and Processing, Wuhan University of Technology, Wuhan, China}

\author{Akio Asami}
\affiliation{Department of Applied Physics and Physico-Informatics, Keio University, Yokohama 223-8522, Japan}

\author{Arne Brataas}
\affiliation{Center for Quantum Spintronics, Department of Physics, Norwegian University of Science and Technology, NO-7491 Trondheim, Norway}

\author{Kazuya Ando\footnote[2]{Correspondence and requests for materials should be addressed to ando@appi.keio.ac.jp}}
\affiliation{Department of Applied Physics and Physico-Informatics, Keio University, Yokohama 223-8522, Japan}
\affiliation{Keio Institute of Pure and Applied Science, Keio University, Yokohama 223-8522, Japan}
\affiliation{Center for Spintronics Research Network, Keio University, Yokohama 223-8522, Japan}

\maketitle

\noindent\textbf{Abstract}

\textbf{Spintronic devices are based on heterojunctions of two materials with different magnetic and electronic properties. Although an energy barrier is naturally formed even at the interface of metallic heterojunctions, its impact on spin transport has been overlooked. Here, using diffusive spin Hall currents, we provide evidence that the inherent energy barrier governs the spin transport even in metallic systems. We find a sizable field-like torque, much larger than the damping-like counterpart, in Ni$_{81}$Fe$_{19}$/Bi$_{0.1}$Sb$_{0.9}$ bilayers. This is a distinct signature of barrier-mediated spin-orbit torques, which is consistent with our theory that predicts a strong modification of the spin mixing conductance induced by the energy barrier. Our results suggest that the spin mixing conductance and the corresponding spin-orbit torques are strongly altered by minimizing the work function difference in the heterostructure. These findings provide a new mechanism to control spin transport and spin torque phenomena by interfacial engineering of metallic heterostructures.
}

\bigskip\noindent
\textbf{Introduction}

Junctions formed at the contact between two different materials are fundamental building blocks of solid-state devices. Prominent examples include junctions consisting of two materials with different electronic properties, such as metal/semiconductor junctions. When a metal is brought into contact with a semiconductor, a potential barrier for electrons is formed, as shown in Fig.~1a, due to the mismatch of the work functions of the two materials~\cite{schottky1939theory,mott1939theory}. Since the energy barrier, known as a Schottky barrier, governs charge transport across the interface, energy barrier engineering plays a key role in improving the performance of electronic devices in a variety of modern electronic applications~\cite{razavieh2019challenges}. 

An alternative technology that exploits the electron spin rather than its charge, spintronics, relies on junctions consisting of two materials with different magnetic properties~\cite{Zutic,RevModPhys.89.025006}. 
Emerging spin-orbitronic technology is based on the ability to manipulate the magnetization in magnetic/nonmagnetic-material junctions through bulk and interfacial spin-orbit couplings~\cite{manchon2019current}. In a magnetic heterostructure, a charge current applied parallel to the interface generates a nonequilibrium spin current and spin polarization via the spin Hall effect due to the bulk spin-orbit coupling in the heavy metal layer~\cite{Dyakonov} and/or the Rashba-Edelstein effect due to the spin-orbit coupling at the interface~\cite{bychkov1984properties,edelstein1990spin,PhysRevB.92.014402}. The nonequilibrium spins arising from the spin-orbit coupling exert spin-orbit torques (SOTs) on the spins in the ferromagnetic layer, enabling electrical manipulation of the magnetization. Manipulation of magnetization by SOTs, including field-like (FL) and damping-like (DL) torques, is responsible for the development of many ultralow-power and fast spin-orbitronic nanodevices based on spin-orbit coupling, such as nonvolatile magnetic memories~\cite{liu2012spin}, nanoscale microwave or terahertz sources~\cite{demidov2012magnetic,kampfrath2013terahertz}, and neuromorphic computing devices~\cite{borders2016analogue}.

Analogous to metal/semiconductor junctions, energy barriers are ubiquitous in solid-state devices, including spin-orbitronic devices. Even in a metal/metal heterojunction, a difference in the work functions of the two adjacent layers results in the formation of an energy barrier, manifested by a contact potential difference. This suggests that interface engineering of metallic heterostructures may provide a route to control and improve the performance of spin-orbitronic devices because the SOTs originating from the bulk spin Hall effect are governed by spin transport across the interface~\cite{brataas2000finite,tserkovnyak2005nonlocal}. Although the impact of an insertion layer on spin transport and SOTs has been investigated extensively~\cite{PhysRevLett.113.097202,PhysRevApplied.11.044070,cui2019field,PhysRevLett.116.186601,moriyama2015anti,li2019modulation}, the role of inherent barriers in metallic spin-orbitronic devices has been overlooked.

In this work, we report the observation of SOTs governed by energy barriers inherent at metallic interfaces. By measuring SOTs for Ni$_{81}$Fe$_{19}$/Bi$_{0.1}$Sb$_{0.9}$ bilayers without any insertion layer, we find that the spin Hall effect in the semimetal Bi$_{0.1}$Sb$_{0.9}$ layer generates a sizable FL torque, which is several times larger than the induced DL torque, in sharp contrast to the conventional SOTs originating from the bulk spin Hall effect in metallic heterostructures. Our observation shows a counterintuitive larger imaginary part $\text{Im}[G^{\uparrow \downarrow}]$ relative to the real part $\text{Re}[G^{\uparrow \downarrow}]$ of the spin mixing conductance in the Ni$_{81}$Fe$_{19}$/Bi$_{0.1}$Sb$_{0.9}$ bilayers, which is consistent with our theory. The theory reveals that spin-dependent hopping and spin-independent hopping determine the spin mixing conductance, which characterizes the spin transmission and reflection, and hence DL and FL torques in ferromagnetic-metal/semimetal junctions. By replacing the Ni$_{81}$Fe$_{19}$ layer with Co$_{72}$Tb$_{28}$ to minimize the size of the energy barrier, we show that spin transmission across the interface becomes more efficient, as evidenced by a more than two orders of magnitude enhancement of $\text{Re}[G^{\uparrow \downarrow}]$, leading to a large DL torque efficiency (up to 0.51) generated by the spin Hall effect in Bi$_{0.1}$Sb$_{0.9}$. These experimental and theoretical results demonstrate that engineering of inherent energy barriers in metallic spintronic devices provide a route to tailor the SOTs torques.

We choose the topological semimetal Bi$_{0.1}$Sb$_{0.9}$ as the source of a diffusive spin current for two reasons. First, the formation of a metal/semimetal contact amplifies the effects of an interfacial energy barrier on spin transport compared to conventional metallic spin-orbitronic devices. The reason for this is that the carrier density of Bi$_{x}$Sb$_{1-x}$ alloys is at least an order of magnitude smaller than that of frequently used heavy metals, such as Pt, but much larger than that of semiconductors; the formation of a metal/semimetal contact results in an atomically thin carrier depletion layer on the semimetal side, as shown in Fig.~1b, which is prominent compared to that in metal/metal heterojunctions. Second, because of the presence of ``hot spots" for Berry curvature in the Brillouin zone, the intrinsic spin Hall conductivity of Bi$_{x}$Sb$_{1-x}$ has been predicted to be comparable to that of heavy metals and much larger than that of semiconductors~\cite{csahin2015tunable}. Along with a well-defined surface in Sb-rich films, these features make semimetallic Bi$_{0.1}$Sb$_{0.9}$ a promising source of a bulk spin current (see also Supplementary Note 1).

\vspace{12pt}
\bigskip\noindent
\textbf{Results}\\
\noindent 
\textbf{SOTs generated by spin Hall effect}

We first investigate the spin transport across a Ni$_{81}$Fe$_{19}$/Bi$_{0.1}$Sb$_{0.9}$ interface through SOT characterization combined with drift-diffusion analysis. Ni$_{81}$Fe$_{19}$(6~nm)/Bi$_{0.1}$Sb$_{0.9}$($t$) bilayers are fabricated by radio frequency (RF) magnetron sputtering, where the numbers in parentheses represent the thickness $t$ (see Methods for details). In Fig.~2a, we show cross-sectional transmission electron microscopy (TEM) images of the Ni$_{81}$Fe$_{19}$/Bi$_{0.1}$Sb$_{0.9}$ bilayer. The TEM images show grains with different crystallographic orientations in the Bi$_{0.1}$Sb$_{0.9}$ layer and a distinct interface between the Ni$_{81}$Fe$_{19}$ and Bi$_{0.1}$Sb$_{0.9}$ layers. The polycrystalline feature of the Bi$_{0.1}$Sb$_{0.9}$ layer with two preferred crystallographic orientations is consistent with the X-ray diffraction results, eliminating any possible variation in the microstructure upon changing the thickness (see Supplementary Note 1).

For the in-plane magnetized Ni$_{81}$Fe$_{19}$/Bi$_{0.1}$Sb$_{0.9}$ films, we measure the spin torque ferromagnetic resonance (ST-FMR) at room temperature, as shown in Fig.~2b. In the device, an in-plane RF current induces a transverse spin current in the Bi$_{0.1}$Sb$_{0.9}$ layer, diffusing towards the Ni$_{81}$Fe$_{19}$ layer. The spin transport across the Ni$_{81}$Fe$_{19}$/Bi$_{0.1}$Sb$_{0.9}$ interface gives rise to oscillating torques with a frequency $f$, including DL and FL torques, exerted on the magnetization of the Ni$_{81}$Fe$_{19}$ layer under the FMR condition. In Fig.~2c, we show representative ST-FMR spectra at different $f$ for the Ni$_{81}$Fe$_{19}$/Bi$_{0.1}$Sb$_{0.9}$ film with $t=10$ nm. The ST-FMR spectra can be decomposed into symmetric ($L_\text{sym}(H)=W^2/[(\mu_0 H-\mu_0H_\text{FMR})^2+W^2]$) and antisymmetric ($L_\text{asym}(H)=W(\mu_0 H-\mu_0H_\text{FMR})/[(\mu_0 H-\mu_0H_\text{FMR})^2+W^2]$) functions~\cite{liu2011spin,fang2011spin}: $V_\text{mix}= V_\text{s} L_\text{sym}(H)+V_\text{a}L_\text{asym}(H)$, where $W$ is the spectral width, $H$ is the external magnetic field, and $H_\mathrm{FMR}$ is the FMR field. As shown in Fig.~2d, we observe a clear symmetric component $V_\text{s}$, determined by the DL effective field $H_\text{DL}$, as well as an antisymmetric component $V_\text{a}$, attributed to the sum of the FL effective field $H_\text{FL}$ and the current-induced Oersted field $H_\text{Oe}$. We determine $H_\text{DL}$ and $H_\text{FL}$ by fitting the ST-FMR spectra at $f=7$~GHz~\cite{gao2018intrinsic,gao2020spin}, where the RF current $I_\mathrm{RF}$ flowing in the device is determined with a vector network analyzer (see Supplementary Note 2). The obtained $H_\text{DL}$ and $H_\text{FL}$ with different Bi$_{0.1}$Sb$_{0.9}$ thicknesses are further converted to the SOT efficiency per unit electric field $E$, $\xi^{E}_\text{DL(FL)}=(2e/\hbar)\mu_{0}M_\text{s}t_\text{F}H_\text{DL(FL)}/E$, as shown in Fig.~3a, where $e$ is the electron charge, $\hbar$ is the reduced Planck constant, $\mu_{0}$ is the vacuum permeability, $M_\text{s}$ is the saturation magnetization, and $t_\text{F}$ is the thickness of the magnetic layer.

Figure~3a shows that both $\xi^{E}_\text{DL}$ and $\xi^{E}_\text{FL}$ monotonically increase over a fairly long length scale with increasing the Bi$_{0.1}$Sb$_{0.9}$ layer thickness $t$. We also note that the magnitude of the FL torque efficiency $\xi^{E}_\text{FL}$ dominates the DL torque efficiency $\xi^{E}_\text{DL}$ at all $t$, which is in sharp contrast to the SOTs in conventional ferromagnetic-metal/heavy-metal devices, where $\xi^{E}_\text{FL}$ is negligible compared to $\xi^{E}_\text{DL}$~\cite{PhysRevB.92.064426}. 
The thickness-dependent SOT efficiencies suggest that the SOTs originate from the bulk spin Hall effect in the Bi$_{0.1}$Sb$_{0.9}$ layer. However, unavoidable self-doping effects and the coexistence of topological surface states (TSSs) and bulk states in the topological semimetal may lead to anomalous thickness-dependent transport properties~\cite{kim2011thickness}, which can subsequently be responsible for the unexpected behavior of thickness-dependent SOT generation. 
To test this possibility, we investigate the transport properties of the Bi$_{0.1}$Sb$_{0.9}$ layers with different $t$. As shown in Fig.~3b, the sheet resistance $R_\text{sh}$ of the Bi$_{0.1}$Sb$_{0.9}$ film is nearly proportional to $1/t$ at room temperature. This result indicates that the transport is dominated by the bulk conduction. In fact, the three-dimensional (3D) resistivity $\rho$ is well reproduced by an empirical model that takes into account the carrier reflection at the surface, as shown in Fig.~3c, demonstrating the bulk-dominated conduction with invariable transport properties upon changing the thickness (see Supplementary Note 3). We also note that the temperature dependence of the conductivity, with an analysis based on a parallel conduction model, suggests the presence of a small hopping term incorporated into the total conduction~\cite{fritzsche1955electrical,gao2012gate}. The nearly thickness-independent contribution of the metallic channel to the total conductivity $\sim 0.93$ reflects the disordered nature but stable transport properties of the sputtered Bi$_{0.1}$Sb$_{0.9}$ films (see Supplementary Note 4).

On the basis of the bulk-dominated conduction with the invariable transport properties, we attribute the thickness-dependent SOT efficiencies to an intrinsic bulk feature of the disordered topological semimetal Bi$_{0.1}$Sb$_{0.9}$. The SOTs induced by the spin Hall effect are described by employing a drift-diffusion approach, where the spin transport at the interface is governed by the complex spin mixing conductance $G^{\uparrow \downarrow}$~\cite{haney2013current,manchon2019current}:
\begin{widetext}
\begin{equation}
\xi^{E}_\text{DL}=\frac{2e}{\hbar}\sigma_\text{s}\left[1-\mathrm{sech}(t/\lambda_\text{s})\right]\frac{2\lambda_\text{s}\rho_\text{BiSb} \text{Re}[G^{\uparrow \downarrow}][2\lambda_\text{s}\rho_\text{BiSb} \text{Re}[G^{\uparrow \downarrow}]+\tanh(t/\lambda_\text{s})]+(2\lambda_\text{s}\rho_\text{BiSb} \text{Im}[G^{\uparrow \downarrow}])^2}
{[2\lambda_\text{s}\rho_\text{BiSb} \text{Re}[G^{\uparrow \downarrow}]+\tanh(t/\lambda_\text{s})]^{2}+(2\lambda_\text{s}\rho_\text{BiSb} \text{Im}[G^{\uparrow \downarrow}])^{2}},
\label{lambdaDL}
\end{equation}
\begin{equation}
\xi^{E}_\text{FL}=-\frac{2e}{\hbar}\sigma_\text{s}\left[1-\mathrm{sech}(t/\lambda_\text{s})\right]\frac{2\lambda_\text{s}\rho_\text{BiSb} \text{Im}[G^{\uparrow \downarrow}]\tanh(t/\lambda_\text{s})}{[2\lambda_\text{s}\rho_\text{BiSb} \text{Re}[G^{\uparrow \downarrow}]+\tanh(t/\lambda_\text{s})]^{2}+(2\lambda_\text{s}\rho_\text{BiSb} \text{Im}[G^{\uparrow \downarrow}])^{2}}.
\label{lambdaFL}
\end{equation}
\end{widetext}
With the extracted Bi$_{0.1}$Sb$_{0.9}$ bulk resistivity, $\rho_\text{BiSb} =350.3$ $\mu$$\Omega$cm, equations~(\ref{lambdaDL}) and (\ref{lambdaFL}) fit the $t$ dependence of $\xi^{E}_\text{DL}$ and $\xi^{E}_\text{FL}$ fairly well, as shown in Fig.~3a, providing the real and imaginary parts of the spin mixing conductance as $\text{Re}[G^{\uparrow \downarrow}]=0.023 \times 10^{13}$ $\Omega$$^{-1}$m$^{-2}$ and $\text{Im}[G^{\uparrow \downarrow}]=0.133 \times  10^{13}$ $\Omega$$^{-1}$m$^{-2}$, respectively. From the fitting, we also obtain the spin diffusion length $\lambda_\text{s} =24.8 \pm 1.0$ nm, which is much longer than that of normal metal SOT source materials, including Pt (2.0 nm)~\cite{PhysRevLett.116.126601}, $\beta$-Ta (2.5 nm)~\cite{allen2015experimental} and W (1.5 nm)~\cite{wang2018comparative}, determined by a similar analysis based on SOT characterization. The bulk spin Hall conductivity in Bi$_{0.1}$Sb$_{0.9}$ is extracted as $\sigma_\text{s} =(3.68 \pm 0.17) \times 10^{5} (\hbar/2e)$ $\Omega$$^{-1}$ m$^{-1}$ (see also Supplementary Note 5).

\vspace{12pt}
\noindent 
\textbf{SOTs governed by inherent energy barrier}

A striking feature of the SOT generation in the Ni$_{81}$Fe$_{19}$/Bi$_{0.1}$Sb$_{0.9}$ bilayers is that the FL torque and $\text{Im}[G^{\uparrow \downarrow}]$ are much larger than the DL torque and $\text{Re}[G^{\uparrow \downarrow}]$, respectively, in sharp contrast to the conventional bulk-dominated SOTs and the spin mixing conductance in metallic devices. It is known that spin transport across a non-magnetic/magnetic interface is well characterized by the spin mixing conductance based on magnetoelectronic circuit theory~\cite{tserkovnyak2005nonlocal}. In this model, the real part of the spin mixing conductance, $\text{Re}[G^{\uparrow \downarrow}]$, determines the spin transmission associated with the DL-torque generation and the imaginary part of the spin mixing conductance, $\text{Im}[G^{\uparrow \downarrow}]$, determines the reflection of spins exchange coupled to the interfacial magnetization associated with the FL-torque generation (see equations~(\ref{lambdaDL}) and (\ref{lambdaFL})). We note that the extracted $\text{Re}[G^{\uparrow \downarrow}]$ is more than three orders of magnitude smaller than that of the Ni$_{81}$Fe$_{19}$/Pt interface~\cite{zhu2019effective}. At intermetallic interfaces, $\text{Re}[G^{\uparrow \downarrow}]$ is close to the Sharvin conductance $G_{\text{Sh}}=(1/\hbar)(ek_\text{F}/2\pi)^2$, which is the limit of the spin mixing conductance, where $k_\text{F}$ is the Fermi wavenumber~\cite{Zwierzycki,PhysRevB.99.045421}. In such system, $\text{Re}[G^{\uparrow \downarrow}]$ is much larger than $\text{Im}[G^{\uparrow \downarrow}]$. For a simple 3D isotropic Fermi surface, $\text{Re}[G^{\uparrow \downarrow}]$ is proportional to $n^{2/3}$ because $k_\text{F}=(3\pi^{2}n)^{1/3}$, where $n$ is the carrier density, suggesting that $\text{Re}[G^{\uparrow \downarrow}]$ is smaller for materials with lower carrier density. However, the large difference in $\text{Re}[G^{\uparrow \downarrow}]$ between the Ni$_{81}$Fe$_{19}$/Bi$_{0.1}$Sb$_{0.9}$ and Ni$_{81}$Fe$_{19}$/Pt interfaces cannot be attributed to the difference in the carrier density between Bi$_{0.1}$Sb$_{0.9}$ and Pt; the carrier density of the disordered Bi$_{0.1}$Sb$_{0.9}$ film is in the range of $0.1 \times 10^{21}~\text{cm}^{-3}<n < 1.2 \times 10^{21}~\text{cm}^{-3}$ (Supplementary Note 6), while that of Pt is $1.6\times 10^{22}$ cm$^{-3}$~\cite{fischer1980mean}, giving a $\text{Re}[G^{\uparrow \downarrow}]$ (Pt)/$\text{Re}[G^{\uparrow \downarrow}]$ (Bi$_{0.1}$Sb$_{0.9}$) ratio of less than 30. Here, the small Re$[G^{\uparrow\downarrow}]$ in the Ni$_{81}$Fe$_{19}$/Bi$_{0.1}$Sb$_{0.9}$ bilayer is also confirmed by measuring the magnetic damping (see also Supplementary Note 7).

The unconventional feature of the SOTs and spin mixing conductance in the Ni$_{81}$Fe$_{19}$/Bi$_{0.1}$Sb$_{0.9}$ heterostructure can be attributed to the formation of an energy barrier at the interface. At the interface of metallic heterostructures, a mismatch of the work functions $\phi$ of the two metals leads to the formation of an interfacial energy barrier, arising from the creation of a carrier depletion region. In the Ni$_{81}$Fe$_{19}$/Bi$_{0.1}$Sb$_{0.9}$ junction, the work function of Ni$_{81}$Fe$_{19}$ is $\phi_{\text{Ni}_{81}\text{Fe}_{19}}= 4.83$~eV~\cite{saito1981work}, while that of $\phi _{\text{Bi}_{0.1}\text{Sb}_{0.9}}$ is estimated to be $\phi _{\text{Bi}_{0.1}\text{Sb}_{0.9}} \sim 4.55$~eV by adopting a two-component alloy model with the work functions of Bi and Sb (Supplementary Note 8). As a result, electron transfer from the Bi$_{0.1}$Sb$_{0.9}$ layer to the Ni$_{81}$Fe$_{19}$ layer appears upon equilibration of the Fermi levels, resulting in the formation of a depletion region of electrons at the atomic scale in the vicinity of the Bi$_{0.1}$Sb$_{0.9}$ side of the Ni$_{81}$Fe$_{19}$/Bi$_{0.1}$Sb$_{0.9}$ interface. 
Since the resulting barrier at the Ni$_{81}$Fe$_{19}$/Bi$_{0.1}$Sb$_{0.9}$ interface is much thinner than that at metal/semiconductor interfaces, no signature of the barrier can be found from current-voltage measurements across the interface. 
Nevertheless, we observed the unconventional feature of the spin mixing conductance; $\text{Im}[G^{\uparrow \downarrow}]$ is much larger than $\text{Re}[G^{\uparrow \downarrow}]$ in the metallic bilayer, which suggests sizable spin-dependent reflection at the interface. This result is in sharp contrast to the conventional wisdom that $\text{Im}[G^{\uparrow \downarrow}]$ is very small in metallic interfaces~\cite{xia2002spin,PhysRevB.70.184438,bauer2005spin,nmat3311,PhysRevApplied.13.054011}. This observation demonstrates that even the ultrathin inherent barrier governs the spin mixing conductance and the corresponding SOT generation, which has been overlooked so far. 
Note that this energy barrier occurs for the transport of both electrons in the conduction band and holes in the valence band from the Bi$_{0.1}$Sb$_{0.9}$ layer to the Ni$_{81}$Fe$_{19}$ layer. Regarding the energy band diagram, a distinct feature of semimetals is that the Fermi level is located above the conduction band minimum and below the valence band maximum (Fig.~1b), rather than pinned in the band gap as in semiconductors (Fig.~1a). Since the Fermi level of Bi$_{0.1}$Sb$_{0.9}$ lies less than $\sim 0.16$ eV above the conduction band maximum at the $L$ point in the Brillouin zone~\cite{xu1993tight}, the interfacial charge transfer gives rise to an electron energy barrier for the spin transport from the Bi$_{0.1}$Sb$_{0.9}$ layer to the Ni$_{81}$Fe$_{19}$ layer. Since the conduction of the Bi$_{0.1}$Sb$_{0.9}$ layer is dominated by hole transport (Supplementary Note 6), taking the hole spin transport across the interface into account is essential. In describing the hole injection from the Bi$_{0.1}$Sb$_{0.9}$ layer to the Ni$_{81}$Fe$_{19}$ layer, or its equivalent electron transport from the Ni$_{81}$Fe$_{19}$ layer to the valence band of the Bi$_{0.1}$Sb$_{0.9}$ layer, one can expect an even higher energy barrier owing to the band bending of the valence band above the Fermi level, as illustrated in Fig.~3d. This corresponds to an energy barrier for the hole spin transport.

To uncover the microscopic origin of the large imaginary part of the spin mixing conductance in the Ni$_{81}$Fe$_{19}$/Bi$_{0.1}$Sb$_{0.9}$ bilayer, we model the depletion region and emergent energy barrier between magnetic-metal and semimetal layers by an interfacial potential barrier consisting of a spin-independent part and a spin-dependent part. Within the microscopic tight-binding formalism, we model the electron hopping through this interfacial potential barrier by a spin-independent hopping amplitude $\delta t$ and a spin-dependent hopping amplitude $J$ (see Fig. \ref{fig4}a). Using a 3D scattering formalism \cite{Tserkovnyak1,Tserkovnyak2,PhysRevLett.128.247204}, we compute the real and imaginary parts of the spin mixing conductance (Supplementary Note 9). Figures \ref{fig4}b and \ref{fig4}c summarize our analytical results. In Fig. \ref{fig4}b, we show that depending on the spin-dependent hopping amplitude $J$, the imaginary part of the spin mixing conductance may be an order of magnitude larger than its real part. Notably, another significant feature is that the interfacial hopping modifies the real part of the spin mixing conductance, which may lead to orders of magnitude enhancement with decreasing size of the energy barrier, as shown in Fig. \ref{fig4}c.

\vspace{12pt}
\noindent 
\textbf{Energy barrier engineering of SOTs}

To verify the generality of the concept of an energy barrier effect on spin transport across metallic heterostructures, we perform a cross check by quantifying the SOT efficiency for perpendicularly magnetized Co$_{72}$Tb$_{28}$(6~nm)/Bi$_{0.1}$Sb$_{0.9}$($t$) films with another widely used harmonic technique. The choice of the Tb-based metallic magnet is due to the low work function of Tb ($3~\mathrm{eV}$), which can minimize the effect of an energy barrier compared to the Ni$_{81}$Fe$_{19}$/Bi$_{0.1}$Sb$_{0.9}$ films. The work function of Co$_{72}$Tb$_{28}$ is $\phi _{\text{Co}_{72}\text{Tb}_{28}}\sim 4.47~\mathrm{eV}$ $(< \phi _{\text{Bi}_{0.1}\text{Sb}_{0.9}}\sim 4.55~\mathrm{eV}) $, and hence there is nearly no energy barrier formed on the Bi$_{0.1}$Sb$_{0.9}$ side, as illustrated in Fig.~3e. 
We employ the out-of-plane angular-dependent harmonic Hall measurement technique to characterize the direction and magnitude of the DL torque for the perpendicularly magnetized device at room temperature (see Fig.~5a)~\cite{yang2020characterization}. 

In Fig. 5b, we show the first and second harmonic Hall resistances, $R_{\omega}$ and $R_{2\omega}$, measured by rotating the external magnetic field in the $xz$ plane. In this result, the contributions from the ordinary Hall effect and the ordinary Nernst effect have been subtracted from the measured first and second harmonic signals, respectively (Supplementary Note 10). We estimate the out-of-plane angle $\theta_{M}$ of the net magnetization at each magnetic field angle $\theta_{H}$ using $\theta_{M}=\arccos(R_{\omega}/R_\text{AHE})-\pi$, where $R_\text{AHE}$ is the anomalous Hall coefficient. Here, the $-\pi$ term arises from the fact that the transport measurement detects the direction of the magnetization of Co because the $4f$ band of Tb is well below the Fermi level, while the net magnetization is dominated by Tb in the Co$_{72}$Tb$_{28}$ film. 
As shown in Fig.~5b, the change in $R_{\omega}$ as a function of $\theta_\text{M}$ opens upwards, which is consistent with the picture of Tb-dominated net magnetization. 
Using $dR_{2\omega}/d(\sin\theta_\text{M})=-R_{2\omega}^\text{ANE}-(1/2)R_\text{AHE}H_\text{DL}[1-4(R_\text{PHE}/R_\text{AHE})^2]/(H+H_\text{K})$, we determine the DL torque efficiency $\xi^{E}_\text{DL}$ for the Co$_{72}$Tb$_{28}$/Bi$_{0.1}$Sb$_{0.9}$ film, where $R_{2\omega}^\text{ANE}$ is the second harmonic signal due to the anomalous Nernst effect, $H_\text{K}$ is the anisotropy field, and $R_\text{PHE}$ is the planar Hall resistance (Supplementary Notes 10 and 11). 
In Fig.~3a, we plot the DL torque efficiency for the Co$_{72}$Tb$_{28}$/Bi$_{0.1}$Sb$_{0.9}$ film. This result provides further evidence of the thickness-dependent $\xi^{E}_\text{DL}$. 
The sign of $\xi^{E}_\text{DL}$ in the Co$_{72}$Tb$_{28}$/Bi$_{0.1}$Sb$_{0.9}$ film is the same as that in the Ni$_{81}$Fe$_{19}$/Bi$_{0.1}$Sb$_{0.9}$ film but opposite to that generated by a Co-Tb single-layer film~\cite{lee2020spin}, supporting that the DL torque originates from the bulk spin Hall effect in the Bi$_{0.1}$Sb$_{0.9}$ layer.

We find that the topological semimetal Bi$_{0.1}$Sb$_{0.9}$ becomes an efficient source of the SOT through energy barrier engineering, i.e., minimizing the mismatch of the work functions of the magnetic layer and the Bi$_{0.1}$Sb$_{0.9}$ layer. Figure 3a shows that the $\xi^{E}_\text{DL}$ of the Co$_{72}$Tb$_{28}$/Bi$_{0.1}$Sb$_{0.9}$ bilayer is more than eight times greater than that of the Ni$_{81}$Fe$_{19}$/Bi$_{0.1}$Sb$_{0.9}$ bilayer when $t=26$ nm. These sizable DL torques are further confirmed by performing magnetization switching on Co$_{72}$Tb$_{28}$/Bi$_{0.1}$Sb$_{0.9}$ with strong perpendicular magnetic anisotropy (see Supplementary Note 12). Note that a DL-torque enhancement due to an exchange torque near the magnetization compensation point, such as that observed in Co$_{1-x}$Gd$_{x}$/Pt~\cite{mishra2017anomalous}, is unlikely to occur in the Co$_{72}$Tb$_{28}$/Bi$_{0.1}$Sb$_{0.9}$ films since the spin Hall angle is nearly unchanged in Co$_{1-x}$Tb$_{x}$/Ta upon changing the Tb concentration ~\cite{finley2016spin}. We also note that the clear difference in $\xi^{E}_\text{DL}$ between the Ni$_{81}$Fe$_{19}$/Bi$_{0.1}$Sb$_{0.9}$ and the Co$_{72}$Tb$_{28}$/Bi$_{0.1}$Sb$_{0.9}$ films cannot primarily be attributed to a potential discrepancy in the estimated efficiencies produced by the different characterization techniques~\cite{gao2020spin}.

The enhancement of the DL torque efficiency is more likely associated with an enhancement in $\text{Re}[G^{\uparrow \downarrow}]$. We obtain $\text{Re}[G^{\uparrow \downarrow}]= 7.11\times 10^{13}$ $\Omega$$^{-1}$m$^{-2}$ at the Co$_{72}$Tb$_{28}$/Bi$_{0.1}$Sb$_{0.9}$ interface by fitting the $t$ dependence of $\xi_\mathrm{DL}^E$ for the Co$_{72}$Tb$_{28}$/Bi$_{0.1}$Sb$_{0.9}$ bilayer using equation~(\ref{lambdaDL}) with the $\sigma_\text{s}$ and {$\lambda_\text{s}$$\rho_\text{BiSb}$} values obtained from the results for the Ni$_{81}$Fe$_{19}$/Bi$_{0.1}$Sb$_{0.9}$ device and under the assumption of Im[$G^{\uparrow \downarrow}$]$\ll${Re}[$G^{\uparrow \downarrow}$] due to the vanished energy barrier (see also Supplementary Note 13), where the dimensionless parameter of the spin-dependent hopping is expected to be in the range of 0.5 to 1. This result suggests that $\text{Re}[G^{\uparrow \downarrow}]$ is enhanced by more than two orders of magnitude by replacing Ni$_{81}$Fe$_{19}$ with Co$_{72}$Tb$_{28}$. The estimated value of $\text{Re}[G^{\uparrow \downarrow}]$ at the Co$_{72}$Tb$_{28}$/Bi$_{0.1}$Sb$_{0.9}$ interface is close to the Sharvin conductance of Bi$_{0.1}$Sb$_{0.9}$, which is in the range of $ 3 \times 10^{13}$ $\Omega$$^{-1}$m$^{-2}$ to $17 \times 10^{13}$ $\Omega$$^{-1}$m$^{-2}$. Here, the maximum $\xi^{E}_\text{DL}$ in the Co$_{72}$Tb$_{28}$/Bi$_{0.1}$Sb$_{0.9}$ film corresponds to the dimensionless DL torque efficiency, or the effective spin Hall angle, of $\xi_\text{DL}=\xi^{E}_\text{DL}\rho = 0.51\pm 0.18$ (see also Supplementary Note 1). This value is larger than that in SOT devices with widely used heavy metals, such as Pt and Ta, and is comparable to that in a SOT device with a recently proposed heavy metal alloy, Au$_{0.25}$Pt$_{0.75}$, with $\xi_\text{DL}=0.35$~\cite{zhu2018highly} (see Supplementary Table I).

To further capture the characteristics of the SOTs originating from the spin Hall effect in topological semimetals, we measure temperature $T$ dependence of the DL torque efficiency $\xi^{E}_\text{DL}$ for the Co$_{72}$Tb$_{28}$(6~nm)/Bi$_{0.1}$Sb$_{0.9}$(10~nm) film, where the interfacial barrier can be neglected. As shown in Fig.~5c, $\xi^{E}_\text{DL}$ is almost independent of $T$ in the Co$_{72}$Tb$_{28}$/Bi$_{0.1}$Sb$_{0.9}$ bilayer. This result is different from the $T$ dependence of the DL torque efficiency in heterostructures consisted of Bi/Sb multilayers (Bi$_x$Sb$_{1-x}$) and CoFeB, where the DL torque efficiency has been found to be suppressed at low $T$~\cite{chi2020spin}.
We note that the spin transport in the Bi$_x$Sb$_{1-x}$/CoFeB film can be affected by an interfacial barrier. In the Bi$_x$Sb$_{1-x}$/CoFeB films, the CoFeB layer is interfaced with Bi. Since the work function of Bi, 4.34 eV, is smaller than that of CoFeB, 4.83 eV, an energy barrier is formed at the Bi$_x$Sb$_{1-x}$/CoFeB interface. The observed $T$-dependent behavior in the Bi$_x$Sb$_{1-x}$/CoFeB films can be understood in terms of spin transport modulated by the thermionic emission; in the presence of the energy barrier, the thermal activation of electrons plays an important role in the electron transport across the interface, which subsequently modifies the spin transport and torque generation efficiency. The suppression of the torque efficiency at low $T$ in the Bi$_x$Sb$_{1-x}$/CoFeB structure might be highly related to the suppression of the spin transmission due to the thermionic emission. In contrast to the Bi$_x$Sb$_{1-x}$/CoFeB structure, no energy barrier is expected to be present at the Co$_{72}$Tb$_{28}$/Bi$_{0.1}$Sb$_{0.9}$ interface, and thus this mechanism is absent in the Co$_{72}$Tb$_{28}$/Bi$_{0.1}$Sb$_{0.9}$ bilayer.

The nearly $T$-independent $\xi^{E}_\text{DL}$ observed for the Co$_{72}$Tb$_{28}$/Bi$_{0.1}$Sb$_{0.9}$ bilayer is consistent with the torque generation originating from the bulk spin Hall effect and spin transmission free from an energy barrier. Considering the fact that the bulk resistivity of the Bi$_{0.1}$Sb$_{0.9}$ layer (350.3~$\mu\Omega$m) falls in the dirty metallic regime, the spin Hall conductivity or $\xi^{E}_\text{DL}$ should scale with the electrical conductivity of the Bi$_{0.1}$Sb$_{0.9}$ layer. However, we found that a change in the conductivity of the Bi$_{0.1}$Sb$_{0.9}$ layer is only less than 9\% upon lowering $T$ from 300 K to 20 K (see Supplementary Note 4). This result indicates that the spin Hall conductivity barely changes with $T$, which is consistent with the nearly $T$-independent $\xi^{E}_\text{DL}$, shown in Fig. 5c. While nontrivial surface states may also be present in the Bi$_{0.1}$Sb$_{0.9}$ layer (see Supplementary Note 14), the $T$-independent behaviour eliminates the possible contribution from the surface states to the DL torque because this contribution is expected to be enhanced at low $T$~\cite{deorani2014observation,wang2015topological}. The minor role of the surface states in generating the DL torque is consistent with the strong Bi$_{0.1}$Sb$_{0.9}$-thickness dependence of the DL torque efficiency, shown in Fig.~3, which evidences the bulk-dominated DL torque generation.

\vspace{12pt}
\noindent 
\textbf{Summary and outlook}

In this work, we have presented the concept of the SOTs governed by an inherent energy barrier in metallic contacts. Experimentally, by measuring the ST-FMR, we found that the DL and FL torque efficiencies in the Ni$_{81}$Fe$_{19}$/Bi$_{0.1}$Sb$_{0.9}$ bilayer increase with the Bi$_{0.1}$Sb$_{0.9}$ thickness, indicating that both SOTs originate from the bulk spin Hall effect. This finding allows us to analyze the Bi$_{0.1}$Sb$_{0.9}$-thickness dependence of the DL and FL torque efficiencies based on the drift-diffusion approach where the interfacial spin transfer is characterized by the spin mixing conductance. This analysis enables us to extract the real and imaginary parts of the spin mixing conductance as Re[$G^{\uparrow\downarrow}$] = 0.023$\times10^{13}$ $\Omega^{-1}$m$^{-2}$ and Im[$G^{\uparrow\downarrow}$] = 0.133$\times10^{13}$ $\Omega^{-1}$m$^{-2}$, respectively. The extracted real part of the spin mixing conductance is two-orders of magnitude smaller than the Sharvin conductance of Bi$_{0.1}$Sb$_{0.9}$. This finding supports the existence of an interfacial energy barrier, which is consistent with the analysis of the work function mismatch. The crucial role of the interfacial barrier in generating the SOTs is further supported by our calculation.As a cross-check, we also aimed to quantify the SOT efficiencies in a Bi$_{0.1}$Sb$_{0.9}$-based system without an interfacial energy barrier. From our analysis of work function mismatch, we found that the Co$_{72}$Tb$_{28}$/Bi$_{0.1}$Sb$_{0.9}$ bilayer meets this requirement. However, the large magnetic damping in Co$_{72}$Tb$_{28}$ makes it difficult to quantify the SOT efficiencies for the Co$_{72}$Tb$_{28}$/Bi$_{0.1}$Sb$_{0.9}$ bilayer by the ST-FMR. Therefore, instead of ST-FMR, we used the second harmonic technique, which is widely considered to be the most reliable for characterizing the SOT efficiency of perpendicularly magnetized films. By taking into account all possible contributions, such as Nernst signals, we extracted the DL-SOT efficiency for the Co$_{72}$Tb$_{28}$/Bi$_{0.1}$Sb$_{0.9}$ bilayer. The result shows that the DL-SOT efficiency of the Co$_{72}$Tb$_{28}$/Bi$_{0.1}$Sb$_{0.9}$ bilayer is clearly higher than that of the Ni$_{81}$Fe$_{19}$/Bi$_{0.1}$Sb$_{0.9}$ bilayer. The corresponding real part of the spin mixing conductance, Re[$G^{\uparrow\downarrow}$] = 7.11$\times10^{13}$ $\Omega^{-1}$m$^{-2}$, of the Co$_{72}$Tb$_{28}$/Bi$_{0.1}$Sb$_{0.9}$ bilayer is found to be close to the Sharvin conductance, as expected from the minor role of the interfacial barrier due to the lower work function of Co$_{72}$Tb$_{28}$.

The above experimental and theoretical observations support the scenario that an inherent barrier governs the SOT generation even in metallic contacts. However, the use of the two different experimental techniques for the SOT characterization could potentially lead to a large discrepancy between the estimated torque efficiencies if artifact contributions dominate the detected signals. To avoid this risk, we have carefully designed our experiments to clarify the role of an interfacial energy barrier in the generation of the SOTs. This includes the comparison of the FL and DL torque efficiencies, the evaluation of the spin mixing conductance, the analysis of work function mismatch, and the development of the theoretical model. We have carefully checked the validity of the ST-FMR analysis (see Supplementary Note 2). We also note that the tiny Re[$G^{\uparrow\downarrow}$] suggested by the SOTs in the Ni$_{81}$Fe$_{19}$/Bi$_{0.1}$Sb$_{0.9}$ bilayer is supported by the magnetic damping measurement (see Supplementary Note 7). We believe that these results provide evidence for the essential role of the inherent barrier in the generation of the SOTs. Here, in this work, we focus on the Co$_{72}$Tb$_{28}$/Bi$_{0.1}$Sb$_{0.9}$ bilayer for the temperature-dependent measurement. At the Ni$_{81}$Fe$_{19}$/Bi$_{0.1}$Sb$_{0.9}$ interface, the presence of finite roughness and stacking faults can introduce atomic inhomogeneity, potentially leading to variations in the energy barrier. Consequently, the temperature-dependent behavior of spin transport across the interfacial energy barrier may be more complex than a simple model of thermionic emission~\cite{tung1991electron}. Further systematic studies are required to fully understand the SOT generation in the presence of inherent barriers, which presents an interesting topic for future research.

The SOTs observed in Ni$_{81}$Fe$_{19}$/Bi$_{0.1}$Sb$_{0.9}$ bilayer are characterized by the large $\text{Im}[G^{\uparrow \downarrow}]$/$\text{Re}[G^{\uparrow \downarrow}]$ ratio even though their origin is the bulk spin Hall effect. This character is different from the widely studied SOTs associated with interfacial spin-orbit coupling in metallic systems~\cite{zhang2015role}. In the latter scenario, the interfacial spin-flip scattering modulates the spin transmission probability across the interface, where the discontinuity of the spin current depends on the strength of the interfacial spin-orbit coupling. This effect, known as spin memory loss, is manifested as a vanishingly small $\text{Im}[G^{\uparrow \downarrow}]$ compared to $\text{Re}[G^{\uparrow \downarrow}]$. Different from this scenario, our experimental and theoretical results show that an energy barrier due to the formation of a carrier depletion region governs the spin mixing conductance by modifying the amplitudes of the spin-dependent and spin-independent hopping. This subsequently has a significant impact on the generation of FL and DL torques.

The role of interfacial barriers in electrical spin injection in ferromagnetic-metal/semiconductor junctions has been discussed over the past two decades. The role of interfacial barriers in the SOT generation by the spin Hall effect is clearly different from that in the electrical spin injection into semiconductors.
In the electrical spin injection into a semiconductor, an electric voltage is applied across a ferromagnetic-metal/semiconductor interface. An interfacial barrier, including a Schottky barrier and an artificial insertion layer, has been used to improve the spin transmission efficiency across the interface because it allows to circumvent the impedance mismatch problem~\cite{Zutic}. In contrast, in the SOT generation by the spin Hall effect, spin-dependent reflection is promoted by an inherent barrier, resulting in the sizable FL torque in the metallic bilayers. It is important to note that due to the extremely low spin transmission probability across Schottky or oxide-insertion layers in the diffusive regime, it has been challenging to investigate the generation of SOTs through a combination of experimental and theoretical studies. Our results demonstrate that an ultrathin energy barrier, naturally formed at the magnetic-metal/semimetal interface, significantly impacts the spin mixing conductance. This provides a unique opportunity to uncover the physics of the generation of SOTs in metallic heterostructures.

Our experimental and theoretical results demonstrate that energy barriers inherent in metallic spintronic devices are the key to control the FL and DL torques. The high carrier density of heavy metals limits the efficient control of the spin Hall effect and SOTs. The advantages of semimetals, such as Bi$_{0.1}$Sb$_{0.9}$, in terms of their relatively low carrier density with sizable SOT generation efficiencies allow to create ultrathin energy barriers, providing a possible route to simultaneously tailor the FL and the DL torques. Of particular importance is that the results open a route to tailor the FL torque by interface engineering. Recent advances in spin-orbitronics have demonstrated the crucial role of the FL torque in spin-orbitronic phenomena and functionalities. In particular, the FL torque dynamically influences the magnetic domain wall chirality, which is a key element for improving the SOT switching efficiency~\cite{baumgartner2017spatially} and realizing unipolar magnetization switching~\cite{lee2018oscillatory}. More recently, a large FL torque is predicted to be crucial for ultrafast and highly efficient field-free switching~\cite{vlasov2022optimal}.

 The formation of an interfacial energy barrier in semimetal-based heterojunctions is a general effect when the Fermi level of the semimetal lies above that of the metal. Given the growing interest in utilizing novel materials such as topological semimetals, including Dirac and Weyl semimetals~\cite{doi:10.1146/annurev-conmatphys-031016-025458,RevModPhys.93.025002}, and metallic van der Waals ferromagnets~\cite{deng2018gate,electricalgating}, our results highlight the crucial role of interfacial energy barriers in the pursuit of high-performance spintronic devices. Moreover, the impact of inherent barriers on the spin mixing conductance is not only important for the generation of SOTs but also for a variety of phenomena induced by spin transport, such as spin Hall magnetoresistance, spin Seebeck effect, and spin pumping. Therefore, our results shed light on the importance of inherent energy barriers, ubiquitous in heterojunctions, in the engineering of spintronic devices and understanding of spintronic phenomena.

\clearpage

\bigskip\noindent
\textbf{Methods}\\
\noindent \textbf{Device fabrication.} 
All  films were grown on thermally oxidized Si substrates at room temperature by RF magnetron sputtering. A Bi$_{0.1}$Sb$_{0.9}$ layer with a thickness $t$ ranging from 6 to 70 nm was first deposited using a composite Bi$_{0.1}$Sb$_{0.9}$ target with the base pressure of the chamber less than $1\times 10^{-5}$ Pa. The composition of the Bi$_{0.1}$Sb$_{0.9}$ films was confirmed by energy-dispersive X-ray spectroscopy. The Ar pressure during the deposition was fixed at 0.25~Pa, and an RF power of 70~W yielded a sputtering rate of 3.2 \AA/s. 
For the transport measurements, the Bi$_{0.1}$Sb$_{0.9}$ single-layer films were patterned into Hall bars with a width of 250 $\mu$m and a length of 1050 $\mu$m using the shadow mask technique. The film surface was capped by 2.5-nm-thick Al$_{2}$O$_{3}$ to prevent oxidation. 
For the devices used for ST-FMR measurements, the Ni$_{81}$Fe$_{19}$ layer was grown on the Bi$_{0.1}$Sb$_{0.9}$ layer at a deposition rate of 0.5 \AA/s, followed by a 2.5 nm-thick Al$_{2}$O$_{3}$ capping layer. The bilayer films were patterned into rectangular strips with a width of 7~$\mu$m and a length of 49~$\mu$m by photolithography and lift-off techniques. 
For the devices used for second harmonic measurements, the Co$_{72}$Tb$_{28}$ layer was prepared by cosputtering Co and Tb targets directly on top of the Bi$_{0.1}$Sb$_{0.9}$ layer and subsequently capped by 3-nm-thick Al to prevent oxidation. The RF power for the Co target was 120~W, and that for the Tb target was 49~W. The composition of the films was determined from the well-calibrated deposition rate. For the second harmonic measurements, the films were patterned into Hall cross bar devices with a width of 10 $\mu$m and a length of 40 $\mu$m using photolithography with a negative resist followed by Ar-ion milling and lift-off techniques.

\bigskip\noindent \textbf{Spin-torque ferromagnetic resonance.}
The SOTs of the in-plane magnetized Ni$_{81}$Fe$_{19}$/Bi$_{0.1}$Sb$_{0.9}$ bilayer films were characterized by ST-FMR. For the measurement, an RF current was applied to the device along its longitudinal direction using an analogue signal generator. The rectified voltage generated from the mixing of the RF current and the oscillating resistance due to the magnetization precession of the Ni$_{81}$Fe$_{19}$ layer was detected on the inductive side of the bias tee with a nanovoltmeter. 
To measure the RF current flowing in the device, the transmission and reflection coefficients were determined using a vector network analyser in the relevant frequency range (4-10 GHz). 

\bigskip\noindent \textbf{Second harmonic measurements.}
For the perpendicular magnetized Co$_{72}$Tb$_{28}$/Bi$_{0.1}$Sb$_{0.9}$ bilayer films, we employed the out-of-plane angular-dependent harmonic Hall measurement technique to characterize the SOTs. This method allows contributions from the anomalous Nernst effect and the ordinary Nernst effect to the second harmonic voltage to be eliminated. The latter can dominate the signal, leading to overestimation of the spin Hall conductivity in topological insulators and semimetals. During the measurement, an alternating current with a frequency of 85 Hz was injected into the device using a multifunction generator. 
The magnetic field slightly deviated from the film normal direction and was rotated in the out-of-plane direction of the sample, where the net magnetization coherently rotated with the external field and no magnetic multidomains were formed. The first and second harmonic Hall voltages were simultaneously detected using two lock-in amplifiers triggered at the same frequency by a current source.

\bigskip\noindent
\noindent\textbf{Data availability}\\
The data that support the findings of this study are available from the corresponding author upon reasonable request.

\clearpage

\bigskip\noindent
\textbf{Acknowledgements}\\
This work was supported by JSPS KAKENHI (Grant Numbers 22H04964, 19H00864, 20H00337, 20H02593), the JST FOREST Program (Grant Number JPMJFR2032), the Canon Foundation, the Asahi Glass Foundation, the JGC-S Scholarship Foundation, the Spintronics Research Network of Japan (Spin-RNJ), and MEXT Initiative to Establish Next-generation Novel Integrated Circuits Centers (X-NICS) (Grant Number: JPJ011438). T.G. acknowledge the financial support by JSPS KAKENHI (Grant Number 22K14561). A.Q., R.E.T., and A.B. were supported by the Research Council of Norway through its Centres of Excellence funding scheme, Project No. 262633, ``QuSpin''.
A.Q. was supported by the Norwegian Financial Mechanism Project No. 2019/34/H/ST3/00515, ``2Dtronics''.

\bigskip\noindent
\textbf{Author contributions}\\
T.G., S.Z., and H.N. fabricated devices. T.G., and S.H. collected and analysed the data with the help of Y.T., and A.A., H.A., and R.T. performed structural characterizations. K.A. and T.G. designed the experiments and developed the explanation. A.Q., R.E.T., and A.B. performed the theoretical calculations and developed the explanation of the experimental results.
T.G. and K.A. wrote the manuscript with the help of A.Q., R.E.T., and A.B. All authors discussed the results and reviewed the manuscript.

\bigskip\noindent
\textbf{Competing interests}\\
The authors declare no competing interests.

\clearpage

\begin{figure}[bt]
\includegraphics[scale=1]{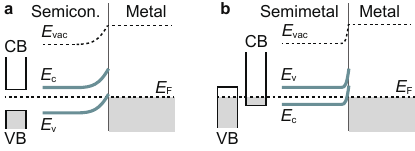}
\caption{\textbf{Energy barrier for charge transport across interfaces.} Illustration of energy band diagrams for \textbf{a,} metal/semiconductor, and \textbf{b}, metal/semimetal contacts in thermal equilibrium. When electrons are transferred from the semiconductor or the semimetal to the metal upon equilibration of the Fermi levels $E_\text{F}$, the semiconductor or semimetal conduction band (CB) edge $E_\text{c}$ and valence band (VB) edge $E_\text{v}$ will bend upwards, leading to the creation of a depletion region. $E_\text{vac}$ represents the vacuum level.}
\label{fig1}
\end{figure}

\clearpage

\begin{figure}[bt]
\includegraphics[scale=1]{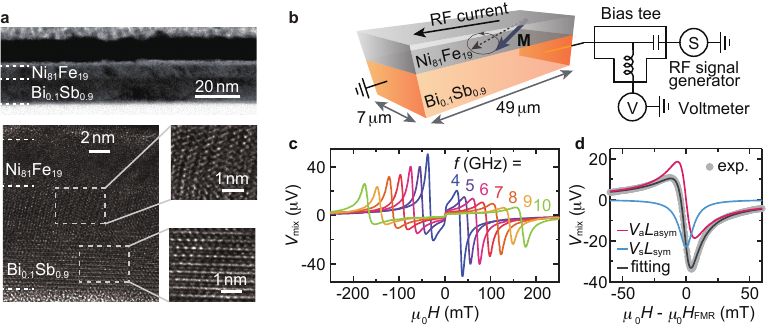}
\caption{\textbf{Structural characterization and ST-FMR.} \textbf{a,} Bright-field cross-sectional TEM image (upper panel) and corresponding high-resolution image (bottom left panel) of the Ni$_{81}$Fe$_{19}$(6 nm)/Bi$_{0.1}$Sb$_{0.9}$(10 nm) bilayer. {The bottom right panel shows enlarged images of the selected regions indicated by the white dashed rectangles in the Bi$_{0.1}$Sb$_{0.9}$ layer.} \textbf{b,} Schematic of the ST-FMR measurement on the in-plane magnetized Ni$_{81}$Fe$_{19}$/Bi$_{0.1}$Sb$_{0.9}$ bilayer. The applied RF current induces effective fields, which drive precession of the magnetization $\bf{M}$ around the external magnetic field $\bf{H}$ (dashed arrow), where $\bf{H}$ is applied at 45$^{\circ}$ with respect to the longitudinal direction of the device. \textbf{c,} Magnetic field $H$ dependence of the rectified voltage $V_\text{mix}$ for the Ni$_{81}$Fe$_{19}$(6 nm)/Bi$_{0.1}$Sb$_{0.9}$(10 nm) bilayer measured at room temperature with RF current frequencies of 4-10 GHz. \textbf{d,} Fitting result of the ST-FMR signal for the sample used in (\textbf{c}) at 7~GHz. The turquoise and red curves present the symmetric Lorentzian ($V_\text{s} L_\text{sym}$) and antisymmetric Lorentzian ($V_\text{a}L_\text{asym}$) components, respectively.}
\label{fig2}
\end{figure}

\clearpage

\begin{figure}[bt]
\includegraphics[scale=1]{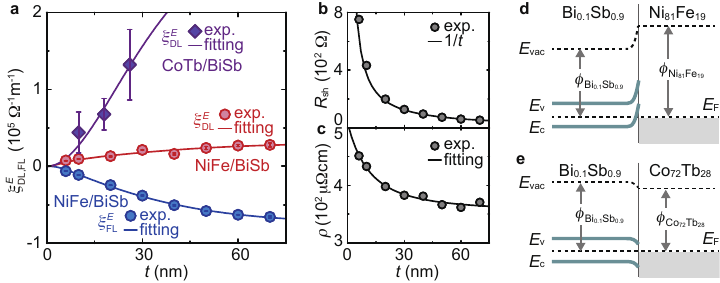}
\caption{\textbf{Characterization of SOTs and transport properties. a,} DL(FL) SOT efficiency per electric field $\xi^{E}_\text{DL(FL)}$ as a function of the Bi$_{0.1}$Sb$_{0.9}$ layer thickness $t$ for the Ni$_{81}$Fe$_{19}$(6 nm)/Bi$_{0.1}$Sb$_{0.9}$ and Co$_{72}$Tb$_{28}$(6 nm)/Bi$_{0.1}$Sb$_{0.9}$ bilayers. The error bars correspond to the standard deviation from the ST-FMR measurements in multiple devices and the linear fit to the second harmonic signals. The solid curves are fits to the $\xi^{E}_\text{DL}$ and $\xi^{E}_\text{FL}$ data obtained by employing the drift-diffusion approach given as equations~(\ref{lambdaDL}) and (\ref{lambdaFL}).
Thickness $t$ dependence of \textbf{b}, sheet resistance $R_\text{sh}$, and \textbf{c}, corresponding 3D resistivity $\rho$ for the Bi$_{0.1}$Sb$_{0.9}$ film. All measurements were performed at room temperature. Simplified energy band diagrams of the \textbf{d}, Ni$_{81}$Fe$_{19}$/Bi$_{0.1}$Sb$_{0.9}$, and \textbf{e}, Co$_{72}$Tb$_{28}$/Bi$_{0.1}$Sb$_{0.9}$ bilayers in thermal equilibrium. At the Ni$_{81}$Fe$_{19}$/Bi$_{0.1}$Sb$_{0.9}$ interface, electron transfer from Bi$_{0.1}$Sb$_{0.9}$ to Ni$_{81}$Fe$_{19}$ occurs to equilibrate the Fermi levels, as the work function $\phi _{\text{Ni}_{81}\text{Fe}_{19}} > \phi _{\text{Bi}_{0.1}\text{Sb}_{0.9}}$. This results in the creation of an electron depletion region at the atomic scale on the Bi$_{0.1}$Sb$_{0.9}$ side. In contrast, at the Co$_{72}$Tb$_{28}$/Bi$_{0.1}$Sb$_{0.9}$ interface, the difference in the work functions gives rise to transfer of electrons from Co$_{72}$Tb$_{28}$ to Bi$_{0.1}$Sb$_{0.9}$ because $\phi _{\text{Co}_{72}\text{Tb}_{28}}< \phi _{\text{Bi}_{0.1}\text{Sb}_{0.9}}$, and thus, nearly no electron barrier is formed on the Bi$_{0.1}$Sb$_{0.9}$ side.} 
\label{fig3}
\end{figure}

\clearpage

\begin{figure}[bt]
\includegraphics[scale=0.85]{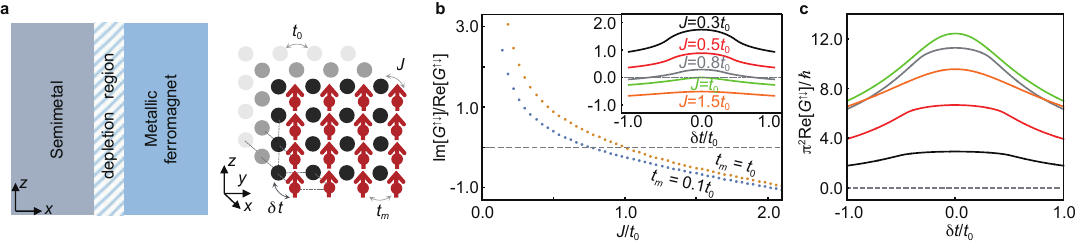}
\caption{\textbf{Theoretical modelling. a,} (Left) Schematic setup of the semimetal/ferromagnetic-metal heterostructure and induced depletion region at the interface. (Right) Semi-infinite 3D cubic lattice structure and its relevant parameters used for the tight-binding model. The ferromagnetic metal is modelled by a 2D layer as a boundary condition with a hopping integral $t_m$, while the semimetal layer is modelled by a semi-infinite region with a hopping integral $t_0$, and $J$ is the exchange coupling between the spin of conduction electrons in the semimetal layer and magnetic moments in the ferromagnetic layer. $J$ and $\delta t=t_0-t_m$ effectively represent spin-dependent and spin-independent hopping across an energy barrier induced by the depletion region. \textbf{b,} Ratio between the imaginary and real parts of the spin mixing conductance $G^{\uparrow\downarrow}$ as a function of spin-independent interfacial hopping amplitude $\delta t$ and spin-dependent interfacial hopping amplitude $J$. \textbf{c,} Real part of the spin mixing conductance Re[$G^{\uparrow\downarrow}$] as a function of $\delta t$ for various $J$.} 
\label{fig4}
\end{figure}

\clearpage

\begin{figure}[bt]
\includegraphics[scale=1]{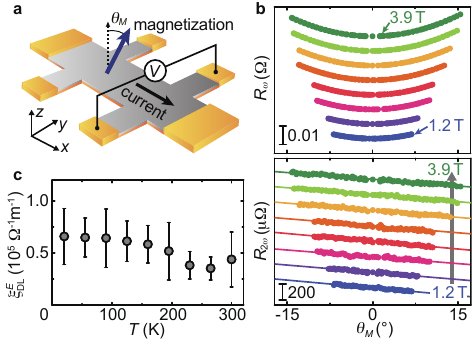}
\caption{\textbf{Harmonic signals and temperature dependence of SOT.} \textbf{a,} Schematic of the out-of-plane angular-dependent harmonic Hall measurement on the perpendicularly magnetized Co$_{72}$Tb$_{28}$/Bi$_{0.1}$Sb$_{0.9}$ bilayer. An alternating current is applied along the longitudinal direction of the device with an external magnetic field $H$ rotating in the $xz$ plane, where the magnetization coherently rotates with $H$. The magnetization angle $\theta_{M}$ represents the angle between the net magnetization and film normal directions. \textbf{b,} First $R_\omega$ and second $R_{2\omega}$ harmonic signals as a function of the magnetization angle $\theta_{M}$ for the Co$_{72}$Tb$_{28}$(6 nm)/Bi$_{0.1}$Sb$_{0.9}$(10 nm) bilayer at room temperature. The signals were recorded simultaneously under different external magnetic fields, 1.2 to 3.9 T. 
\textbf{c,} DL torque efficiency per electric field $\xi^{E}_\text{DL}$ as a function of temperature $T$ for the Co$_{72}$Tb$_{28}$(6 nm)/Bi$_{0.1}$Sb$_{0.9}$(10 nm) bilayer. The error bars denote the standard deviation from the linear fit to the second harmonic signals. 
}

\label{fig5}
\end{figure}

\end{document}